\documentclass[onecolumn,final,runningheads,natbib]{svjour2}
\smartqed  
\usepackage{graphicx}
\journalname{Astrophysics and Space Science}
\begin{document}

\title{Slow glitches in the pulsar B1822$-$09
}


\author{Tatiana V. Shabanova}


\institute{T.V. Shabanova \at Pushchino Radio Astronomy
Observatory, Astro Space Center, P.N. Lebedev
Physical Institute, Russian Academy of Sciences, Pushchino, 142290, Russia\\
\email{tvsh@prao.psn.ru} }

\date{Received: 2006 June 21 / Accepted: 2006 June 30}

\maketitle

\begin{abstract}
The pulsar B1822$-$09 (J1825$-$0935) experienced a series of five
unusual slow glitches over the 1995--2004 interval. The results
of further study of this unusual glitch phenomenon are presented.
It is also reported the detection a new glitch of typical signature
that occurred in the pulsar period in 2006 January.

\end{abstract}

\section{Introduction}
\label{intro} The timing observations of PSR B1822$-$09 obtained
with the Pushchino radio telescope have revealed a new type of
glitches, which has not been observed in any pulsar before. These
rotation variations occurred in the form of slow glitches
(\citealt{sh1998}, \citealt{shu2000}, \citealt{sh2005}, hereafter
SH05). The present paper also reports the detection a new glitch
of small size that occurred in 2006 January.

Characteristic feature of the slow glitches observed is a gradual
exponential increase in the rotation frequency $\nu$ with a
time-scale of 200--300 d. The curvature of the $\Delta\nu$ curve
is determined by the corresponding change in the frequency
derivative $\dot\nu$, the magnitude of which decreases by $\sim$
1--2 $\%$ of the initial value across the glitch. No obvious
relaxation in frequency after a slow glitch is observed. The size
of the slow glitches after a span of a few years is rather
moderate, with magnitude of $\Delta\nu/\nu \sim 2\times10^{-8}$.
Three slow glitches of similar amplitude have occurred in 1995
June, 1998 August and 2000 December. The third slow glitch was
independently observed by \cite{zou2004}. The authors also
reported a fourth smaller slow glitch that occurred in 2003.
\section{Observations}
\label{sec:1} Timing observations of the pulsar were performed
with the BSA transit radio telescope at the Pushchino Observatory
at frequencies around 103 and 112 MHz, using a 32$\times$20 kHz
filter bank receiver, as described in detail in SH05. The
topocentric arrival times for each observation were corrected to
the barycenter of the Solar System using the TEMPO software
package and the JPL DE200 ephemeris. A simple spin-down model
involving $\nu$ and $\dot\nu$ was used for fitting the
barycentric arrival times. In order to study variations in the
spin-down parameters of the pulsar in more detail, $\nu$ and
$\dot\nu$ were calculated by performing local fits to the arrival
time data over the intervals of $\sim$200 d.
\section{Results}
\label{sec:2} 
\begin{figure*}
\centering
\includegraphics[width=0.75\textwidth]{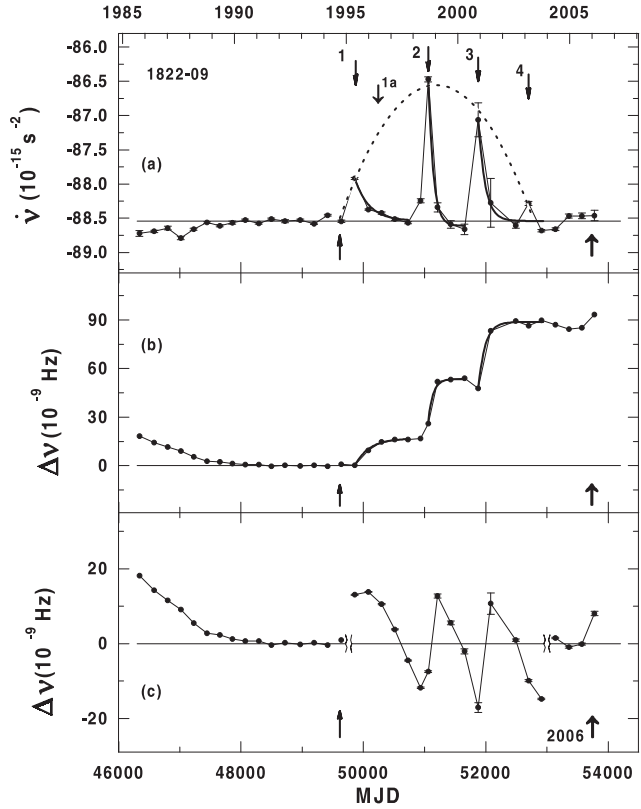}
\caption{$\dot\nu$ and $\Delta\nu$ as a function of time. (a) The
rapid decreases in $\dot\nu$ over the interval 1995--2004 are the
effect of the slow glitches. (b) $\Delta\nu$ relative to a fit to
the data for the interval 1991--1994. (c) As for (b) but with
$\Delta\nu$ relative to a new fit to the data for the 1995--2004
interval where the slow glitches occurred (middle part of the
plot) and for the 2004--2005 interval (right hand side of the
plot). Arrows pointing downwards indicate the epochs at which the
slow glitches occurred while arrows pointing upwards indicate the
normal glitches.}
\label{shaban:1}       
\end{figure*}

With respect to SH05, we extend the observational
interval up to 2006 June, including two years of new observations,
and present a description of the timing behavior of the pulsar
over the 21-yr data span from 1985 to 2006. The timing data set
includes the Pushchino data collected for the period 1991--2006
and the Hartebeesthoek data collected over the 1985--1998 interval
and taken from the previously published paper \citep{shu2000}.
Fig.~\ref{shaban:1} gives the dependencies of $\dot\nu$ and the
frequencies residuals $\Delta\nu$ on time, plotted as in fig.2 of
the paper SH05, but supplemented with the four last new points.

Since 1994 the pulsar underwent a series of glitches. The first
glitch that occurred in 1994 September (MJD 49615) had a typical
signature and an extremely small size with the fractional increase
of the rotational frequency $\sim 8\times10^{-10}$ (marked by the
bottom arrow in Fig.~\ref{shaban:1}). The next glitch occurred
about a year later, in 1995 June, and initiated a series of five
glitches of unusual signature, showing a slow growth in the
frequency rotation during hundreds days. Here we mention five
glitches because the first glitch shown in Fig.~\ref{shaban:1},
in fact, represents the sum of two partially overlapped glitches of
the smaller size (as was pointed out in fig.1 of SH05). The
parameters of the slow glitches are listed in
Table~\ref{shabtab:1}. The epochs of glitches correspond to the
time at which $\dot\nu$ reaches its minimum value. The interval
between all the glitches is approximately equal to 800 d, if the
small glitch~1a is taken as the starting point. With all
probability the small glitch~1a marks the starting point of a new
phase in the glitching behaviour of PSR B1822$-$09. It is seen
that rather small sizes of the glitches $\Delta\nu$ are related to
large changes of $\dot\nu$ across the glitch, which reach $\sim
2\%$. These $\Delta\dot\nu$ are responsible for the steepness of
the front in $\Delta\nu$.

Fig.~\ref{shaban:1}(a) shows that all the peaks of $\Delta\dot\nu$
lie on a curve which is the envelope of these peaks and is well
described by a parabolic curve. The existence of the envelope
indicates that all the slow glitches are the components of one
process, the action of which ceased in the middle of 2004. The
beginning of the envelope coincides with the epoch of the 1994
glitch of typical signature. It is likely that this small glitch
has behaved as a trigger for the following unusual glitch-like
events. Fig.~\ref{shaban:1}(c) also shows that the process,
responsible for the oscillatory changes in the rotation frequency,
was stopped in the middle of 2004.

\begin{figure*}
\centering
\includegraphics[width=0.75\textwidth]{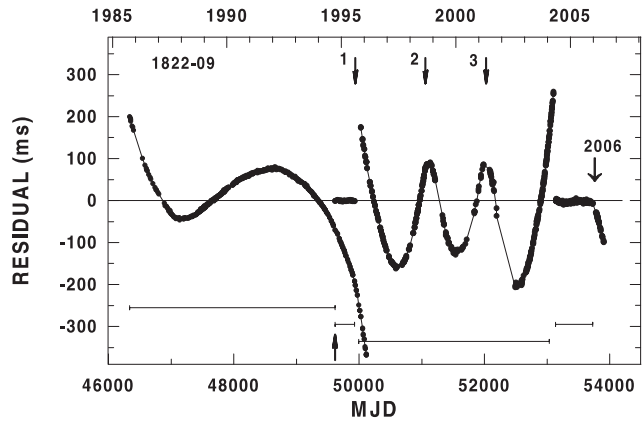}
\caption{Timing residuals of the pulsar over the 21-yr data span
between 1985 and 2006, obtained from four independent fits for
$\nu$ and $\dot\nu$ over four different intervals indicated in the
plot by the horizontal lines. The three upper arrows indicate the
epochs at which the three largest slow glitches occurred. The
bottom arrow indicates the epoch at which the 1994 glitch of
typical signature occurred. The new glitch occurred in January
2006.}
\label{shaban:2}       
\end{figure*}

Fig.~\ref{shaban:2} presents the timing residuals of the pulsar,
plotted as in fig.3 of SH05, but supplemented with the last data
segment from 2004 June to 2006 June.
Analysis of this data segment showed that the pulsar suffered a
new glitch that occurred in 2006 January 10 (MJD 53745(2)). This
glitch is small, with the fractional increase
$\Delta\nu/\nu=(6.6\pm 0.5)\times{10^{-9}}$. The frequency and
timing residuals for this glitch are shown in the rightmost sides
of Fig.~\ref{shaban:1} and Fig.~\ref{shaban:2}.

Thus, a total of seven glitches have been detected in PSR
B1822$-$09 during the 12 years since 1994. All glitches are small,
with fractional increases of the rotation frequency $\Delta\nu/\nu
\sim (0.8-32)\times{10^{-9}}$. Five of these glitches belong to a
new type of glitches that occurred in the form of slow
glitches. The newly reported glitch of 2006, on the contrary, has
a typical signature.

%
%
\begin{table}[t]
\caption{Slow glitches in PSR B1822$-$09}
\label{shabtab:1}       
\begin{tabular}{lllll}
\hline\noalign{\smallskip}
No. & MJD & $\Delta{\nu}/{\nu}(10^{-9})$ &
 $\Delta{\dot{\nu}}/\dot{\nu}(10^{-3})$ \\[3pt]
\tableheadseprule\noalign{\smallskip}
1  & 49857 & 12.8(2)   & 7.0(2)   \\
1a & 50253 & 4.3(2)    & 4.8(3)   \\
2  & 51060 & 28.7(6)   & 24.2(4)  \\
3  & 51879 & 32.0(9)   & 16.7(8)  \\
4  & 52700 & 2.5(3)    & 2.9(3)   \\
\noalign{\smallskip}\hline
\end{tabular}
\end{table}


\end{document}